\documentclass[hidelinks,letterpaper]{IEEEtran}

\usepackage{amsmath}
\usepackage{amsfonts}
\usepackage{amssymb}
\usepackage{amsthm}
\usepackage{tabularx}
\usepackage{mathrsfs} 

\usepackage{url}
\usepackage{hyperref}
\newtheorem{theorem}{Theorem}
\newtheorem{definition}{Definition}

\newtheorem{corollary}{Corollary}

\usepackage{stfloats}
\usepackage{float}
\usepackage{graphicx}
\usepackage{cite}
\usepackage{xcolor}
\renewcommand{\vec}[1]{\mathbf{#1}}

\makeatletter
\def\blfootnote{\xdef\@thefnmark{}\@footnotetext}
\makeatother

\begin{document}
\bstctlcite{IEEEexample:BSTcontrol}
\title{\huge{The Role of Correlation in the Doubly Dirty Fading\\ MAC with Side Information at the Transmitters}}
\author{Farshad~Rostami~Ghadi\IEEEmembership{},
        Ghosheh~Abed~Hodtani\IEEEmembership{},~and  F.~Javier~López-Martínez\IEEEmembership{}}
\maketitle
\begin{abstract}
We investigate the impact of fading correlation on the performance of the doubly dirty fading multiple access channel (MAC) with non-causally known side information at transmitters. Using  Copula theory, we derive closed-form expressions for the outage probability (OP) and the coverage region under positive/negative dependence conditions. We show that a positive dependence structure between the fading channel coefficients is beneficial for the system performance, as it improves the OP and extends the coverage region compared to the case of independent fading. Conversely, a negative dependence structure has a detrimental effect on both performance metrics.
\end{abstract}
\maketitle
\blfootnote{\noindent Manuscript received April xx, 2021; revised XXX. This work has been funded in part by the Spanish Government and the European Fund for Regional Development FEDER (project  TEC2017-87913-R) and by Junta de Andalucia (project P18-RT-3175). The review of this paper was coordinated by XXXX. }

\blfootnote{\noindent F.R. Ghadi and G.A. Hodtani are with Department of Electrical Engineering, Ferdowsi University of Mashhad, Mashhad, Iran. (e-mail: $\rm \{f.rostami.gh,ghodtani\}@gmail.com$).}

\blfootnote{\noindent F.J. Lopez-Martinez is with Departmento de Ingenieria  de Comunicaciones, Universidad de Malaga - Campus de Excelencia Internacional Andalucia Tech., Malaga 29071, Spain  (e-mail: $\rm fjlopezm@ic.uma.es$).}

\blfootnote{Digital Object Identifier 10.1109/XXX.2021.XXXXXXX}
%\IEEEpeerreviewmaketitle
\vspace{-3mm}
\section{Introduction}\label{introduction}
Achieving reliability constraints in applications like connected robotics and autonomous systems \cite{saad2019vision} is a key open challenge in the roadmap to sixth-generation (6G) technology. In this regard, multi-user wireless communications techniques that take advantage of side information (SI) at the transmitters can be of great interest, since such knowledge -- either channel state information (CSI), or interference awareness -- can be leveraged to intelligently encode their information. By doing so, the destructive effects of the interference can be reduced, and reliable communication with higher rates can be achieved.

The use of SI at the transmitter was first studied by Shannon in the context of single-user communication systems \cite{shannon1958channels}. For a multi-user setting, Jafar provided a general capacity region for a discrete and memoryless multiple-access channel (MAC) with causal and non-causal independent SI in \cite{jafar2006capacity}. By exploiting a random binning technique, Philosof$-$Zamir extended Jafar's work and presented achievable rate regions for the discrete and memoryless MAC with correlated SI known non-causally at the encoders \cite{philosof2008technical}. The case of a two-user Gaussian MAC with SI at both transmitters (i.e, doubly dirty MAC) for the high-SNR and strong interference regimes was studied in \cite{philosof2011lattice}, on which the achievable rate regions suffer from a \emph{bottleneck} effect dominated by the weaker user compared to the case of a clean MAC (i.e., without interference). 

In wireless communication theory, dependence structures associated to random phenomena in temporal, frequency or spatial scales are often neglected for the sake of tractability \cite{Biglieri2016}. This is the case, for instance, of multi-user channels, where due to physical proximity of the transmitters the channel coefficients observed by each user are in general not independent. One plausible approach to incorporate arbitrary dependence structures that is recently gaining momentum in the wireless communication arena is the use of Copula theory \cite{nelsen2007introduction,Besser2020copula}. Copulas are widely used in statistics, survival analysis, image processing, and machine learning. Recently, they have also become popular in the context of performance analysis of wireless communication systems, supported by empirical evidences \cite{Peters2014} and motivated by the possibility of designing and intelligently controlling dependence structures to improve system performance \cite{Besser2021}. Specifically, Copulas have been used to provide general bounds on the outage performance for dependent slow-fading channels in \cite{Besser2020copula}; to study the physical layer security performance in a correlated Rayleigh fading wiretap channel in \cite{ghadi2020copula1}, and also in \cite{besser2020bounds}; to analyze the impact of interference correlation in the context of ad-hoc networks \cite{ghadi2020copula2}. Finally, the authors in \cite{ghadi2020copula} derived closed-form expressions for the OP and the coverage region in the correlated Rayleigh fading clean MAC, bringing out the positive effect of a \textit{negative} dependence between fading channels in the system performance. 

Given the relevance of the two-user MAC as a key building block in communication theory \cite{Zhang2021}, we study the impact of fading correlation on the performance of doubly dirty MAC with non-causally known SI at transmitters. Differently from the case on which interferences are not present, our theoretical results show that \textit{positive dependence between the fading channel coefficients is beneficial}, since it allows for reducing the OP and extending the coverage region compared to the baseline case of independent fading.
\vspace{-2mm}
\section{System model and definitions}
\subsection{The wireless doubly dirty MAC}
\begin{figure}[!t]\vspace{0ex}
	\centering
	\includegraphics[width=.6\columnwidth]{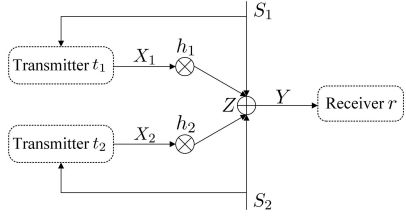} %\textwidth
	\caption{System model depicting a wireless doubly dirty MAC} %\vspace{-0.6cm}
	\label{system-model}
\end{figure}

We consider a two-user wireless doubly dirty MAC \cite{philosof2011lattice} with two known interferences $S_1$ and $S_2$ (see Fig. \ref{system-model}), where, transmitters (users) $t_1$ and $t_2$ send the inputs $X_1$ and $X_2$, respectively. 
%\vspace{0ex}
Therefore, the received signal $Y$ at receiver (base station) $r$ can be defined as:%\vspace{-1ex}
\begin{align}\label{y-signal}
Y=h_1X_1+h_2X_2+S_1+S_2+Z,
\end{align}
where $Z$ represents the Additive White Gaussian Noise (AWGN) with zero mean and variance $N$ (i.e., $Z\sim\mathcal{N}(0,N)$) at the receiver $r$, and $h_1$ and $h_2$ are the corresponding fading channel Rayleigh coefficients, meaning that the channel power gains (i.e., $g_1=|h_1|^2$ and $g_2=|h_2|^2$) are exponentially distributed. We consider the general case on which the fading processes $h_1$ and $h_2$ are correlated. We assume that the interference signals $S_1$ and $S_2$ with variances $Q_1$ ($S_1\sim \mathcal{N}(0,Q_1)$) and $Q_2$ ($S_2\sim \mathcal{N}(0,Q_2)$) are known non-causally at the transmitters $t_1$ and $t_2$, respectively; and the inputs $X_1$ and $X_2$ sent by transmitters $t_1$ and $t_2$ over the channels are subjected to the average power constraint as $\mathbb{E}[|X_1|^2]\leq P_1$ and $\mathbb{E}[|X_2|^2]\leq P_2$, respectively. Besides, we define the signal-to-noise ratio (SNR) at transmitters $t_1$ and $t_2$ as $\gamma_1=\frac{P_1|h_1|^2}{N}$ and $\gamma_2=\frac{P_2|h_2|^2}{N}$, so that the corresponding average SNRs are given by $\bar{\gamma}_1=\frac{P_1\mathbb{E}[|h_1|^2]}{N}$ and $\bar{\gamma}_2=\frac{P_2\mathbb{E}[|h_2|^2]}{N}$, respectively. Therefore, the marginal distributions for the SNR $\gamma_i,\; i=1,2$ are given by $f(\gamma_i)=\frac{e^{-\frac{\gamma_i}{\bar{\gamma}_i}}}{\bar{\gamma}_i}, F(\gamma_i)=1-e^{\frac{-\gamma_i}{\bar{\gamma}_i}}$.
\vspace{-2mm}
\subsection{Preliminary definitions}
\begin{theorem}\label{thm-capacity}
In a block fading doubly dirty MAC with a coherent receiver (fading coefficients $h_1$ and $h_2$ are known at the receiver) and two independent interferences $S_1$ and $S_2$ non-causally known at transmitters $t_1$ and $t_2$, the instantaneous capacity region is determined as follows as long as the interferences $S_1$ and $S_2$ are strong (i.e., $Q_1$, $Q_2 \rightarrow\infty$)\cite{philosof2011lattice}%\vspace{-1ex}
\begin{align}\label{capacity region}
R_1+R_2\leq \tfrac{1}{2}\log_2\left(1+\min\left\{\tfrac{P_1|h_1|^2}{d_1^\alpha N},\tfrac{P_2|h_2|^2}{d_2^\alpha N}\right\}\right),
\end{align}
which is independent of $S_1$ and $S_2$, and where $R_1$ and $R_2$ are the transmission rates for $t_1$
and $t_2$ located at distances $d_1$ and $d_2$, respectively, and $\alpha>2$ is the path loss exponent.
\end{theorem}
We now briefly review some basic definitions and properties of the two-dimensional Copulas \cite{nelsen2007introduction}.
\begin{definition}[Copula]
	Let $\vec{V}=(V_1,V_2)$ be a vector of two random variables (RVs) with marginal cumulative distribution functions (CDFs) $F(v_j)=\Pr(V_j\leq v_j)$ for $j=1,2$, respectively. The relevant bivariate CDF is defined as:%\vspace{-1ex}
\begin{align}
F(v_1,v_2)=\Pr(V_1\leq v_1,V_2\leq v_2).
\end{align}
Then, the Copula function $C(u_1,u_2)$ of $\vec{V}=(V_1,V_2)$ defined on the unit hypercube $[0,1]^2$ with uniformly distributed RVs $U_j:=F(v_j)$ for $j=1,2$ over $[0,1]$ is given by
\begin{align}
C(u_1,u_2)=\Pr(U_1\leq u_1,U_2\leq u_2).
\end{align}
\end{definition}
\begin{theorem}[Sklar’s theorem]
	Let $F(v_1,v_2)$ be a joint CDF of RVs with margins $F(v_j)$ for $j=1,2$. Then, there exists one Copula function $C$ such that for all $v_j$ in the extended real line domain $\bar{R}$,% 
	\vspace{-1ex}
	\begin{align}\label{sklar}
	F(v_1,v_2)=C\left(F(v_1),F(v_2))\right).
	\end{align}
%	\vspace{-2ex}
\end{theorem}
\begin{corollary}\label{col-pdf}
By applying the chain rule to \eqref{sklar}, the joint probability density function (PDF) $f(v_1,v_2)$ is derived as: 	\vspace{-1ex}
\begin{align}
f(v_1,v_2)=f(v_1)f(v_2)c\left(F(v_1),F(v_2)\right),
\end{align}
where $c\left(F(v_1),F(v_2)\right)=\frac{\partial^2 C(F(v_1),F(v_2))}{\partial F(v_1)\partial F(v_2)}$ is the Copula density function and $f(v_j)$ for $j=1,2$ are the marginal PDFs, respectively.
\end{corollary}
\begin{definition}\label{def-surv}
	For a vector of two RVs $\vec{V}=(V_1,V_2)$ with joint CDF $F(v_1,v_2)$ and marginal survival functions $\bar{F}(v_j)=\Pr(V_j>v_j)=1-F(v_j)$ for $j=1,2$, the joint survival function $\bar{F}(v_1,v_2)$ is given by\vspace{-1ex}
	\begin{align}
	&\bar{F}(v_1,v_2)=\Pr(V_1>v_1,V_2>v_2)\\
	&=\bar{F}(v_1)+\bar{F}(v_2)-1+C(1-\bar{F}(v_1),1-\bar{F}(v_2))\\
	&=\hat{C}(\bar{F}(v_1),\bar{F}(v_2)),
	\end{align}
	where $\hat{C}(u_1,u_2)=u_1+u_2-1+C(1-u_1,1-u_2)$ is the survival Copula of $\vec{V}=(V_1,V_2)$.
\end{definition}
The above definitions hold for any arbitrary choice of Copula. We will now exemplify how the outage performance can be characterized in the closed-form expression for the upper/lower Fr\'echet-Hoeffding, Frank, and FGM Copulas. These choices are justified because they capture positive/negative dependences between RVs for any range of correlation, while offering good mathematical tractability. These are also symmetric Copulas, so that their survival Copula coincides with the Copula, i.e., $\hat{C}(u_1,u_2)=C(u_1,u_2).$
\begin{definition}[Fr\'echet-Hoeffding bounds]\label{bounds} For any Copula function $C:[0,1]^2 \mapsto [0,1]$ and any $(u_1, u_2) \in [0,1]^2$, the following bounds hold:
	\begin{align}\label{fbounds}
		C^{-}(u_1,u_2)\leq C(u_1,u_2)\leq C^{+}(u_1,u_2),
	\end{align}
where $C^{-}$ and $C^{+}$ are the lower and upper Fr\'echet-Hoeffding bounds respectively that are defined as follows:
\begin{align}
&C^{-}=\max(u_1+u_2-1,0)\\
&C^{+}=\min(u_1,u_2).
\end{align}
\end{definition}
Note that the lower and upper bounds in \eqref{fbounds} are themselves Copulas and denote the negative and positive perfect correlation, respectively. In addition, the lower and upper Fr\'echet-Hoeffding bounds are symmetric Copulas, i.e.,  $\hat{C}^{-}(u_1,u_2)=C^{-}(u_1,u_2)$ and $\hat{C}^{+}(u_1,u_2)=C^{+}(u_1,u_2)$.
\begin{definition}[Frank Copula]\label{Frank}
The bivariate Frank Copula with dependence parameter $\theta_{\rm fr}\in(-\infty,\infty)\backslash\{0\}$ is defined as:
\begin{align}
C_{\rm fr}(u_1,u_2)=-\tfrac{1}{\theta_{\rm fr}}\ln\left[1+\tfrac{\left(e^{-\theta_{\rm fr} u_1}-1\right)\left(e^{-\theta_{\rm fr} u_2}-1\right)}{e^{-\theta_{\rm fr}}-1}\right],
\end{align}
which accepts both negative and positive dependence structures. When $\theta_{\rm fr}$ tends to $-\infty$ and $+\infty$, the lower and upper bound of  Fr\'echet-Hoeffding will be attained,  respectively. Also, independence is achieved as $\theta_{\rm fr}$ approaches zero.
	\end{definition}
\begin{definition}\label{def}[FGM Copula] The bivariate FGM Copula with dependence parameter $\theta_F\in[-1,1]$ is defined as:
\begin{align}\label{fgm}	\vspace{-2ex}
C_F(u_1,u_2)=u_1u_2\left(1+\theta_F(1-u_1)(1-u_2)\right),
\end{align}
where $\theta_F\in[-1,0)$ and $\theta_F\in(0,1]$ denote the negative and positive dependence structures respectively, while $\theta_F=0$ indicates the independence structure.%\vspace{-3ex}
\end{definition}
\section{Outage probability}
The outage probability is a key metric to evaluate the performance of communication systems operating over fading channels, and is defined as the probability that the channel capacity is less than a certain information rate $R_o>0$, as:%\vspace{-1ex}
\begin{align}
P_{out}&=\Pr(R_1+R_2\leq R_{o})\\
&=\Pr\left(\min\left\{\frac{\gamma_1}{d_1^\alpha},\frac{\gamma_2}{d_2^\alpha}\right\}\leq 2^{2R_o}-1\right)\\
&=1-\Pr\left(\gamma_1>\beta_1,\gamma_2>\beta_2\right)\\
&=1-\hat{C}(\bar{F}_{\gamma_1}(\beta_1),\bar{F}_{\gamma_2}(\beta_2)).\label{surv}
\end{align}
where $\beta_1=d_1^\alpha(2^{2R_o}-1)$ and $\beta_2=d_2^\alpha(2^{2R_o}-1)$.
\begin{theorem}
The OP over correlated Rayleigh fading doubly dirty MAC with defined parameters $\bar{\gamma}_1$, $\bar{\gamma}_2$, $\theta_F$, $\theta_{\rm fr}$, $\beta_1$, and $\beta_2$ is given by \\
(i) as \eqref{out-lower} under lower Fr\'echet-Hoeffding Copula 
\begin{align}\label{out-lower}
	P^{-}_{out}=1-\max\left(e^{-\frac{\beta_1}{\bar{\gamma}_1}}+e^{-\frac{\beta_2}{\bar{\gamma}_2}}-1,0\right)
\end{align}
(ii) as \eqref{out-upper} under upper Fr\'echet-Hoeffding Copula
\begin{align}\label{out-upper}
P^{+}_{out}=1-\min\left(e^{-\frac{\beta_1}{\bar{\gamma}_1}},e^{-\frac{\beta_2}{\bar{\gamma}_2}}\right)
\end{align}
(iii) as \eqref{out-frank} under Frank Copula
\begin{align}\nonumber
&P^{\rm fr}_{out}=\,2-e^{-\frac{\beta_1}{\bar{\gamma}_1}}+e^{-\frac{\beta_2}{\bar{\gamma}_2}}\\
&+\frac{1}{\theta_{\rm fr}}\ln\left[1+\frac{\left(e^{-\theta_{\rm fr} (1-e^{-\frac{\beta_1}{\bar{\gamma}_1}})}-1\right)\left(e^{-\theta_{\rm fr} (1-e^{-\frac{\beta_2}{\bar{\gamma}_2}})}-1\right)}{e^{-\theta_{\rm fr}}-1}\right]\label{out-frank}
\end{align}
(iv) as \eqref{out} under FGM Copula
\begin{align}
\hspace{-1.5ex}P_{out}^F=1-e^{-\left(\frac{\beta_1}{\bar{\gamma}_1}+\frac{\beta_2}{\bar{\gamma}_2}\right)}\left(1+\theta_F\left(1-e^{-\frac{\beta_1}{\bar{\gamma}_1}}\right)\left(1-e^{-\frac{\beta_2}{\bar{\gamma}_2}}\right)\right).\label{out}
\end{align}
\begin{proof}
Using the concept of survival Copula from Definition \ref{def-surv} to Fr\'echet-Hoeffding, Frank, and FGM Copulas, and inserting these survival Copulas into \eqref{surv} the proof is completed.
\end{proof}
\end{theorem}
\begin{figure*}[b]
	\normalsize
	\hrulefill
	\setcounter{equation}{23}
\begin{align}\nonumber\label{coverage region}
		R_1+R_2&\leq\tfrac{\sqrt{\pi}}{2\ln2}\Bigg(
		\tfrac{\bar{\gamma}_2 e^{\frac{d_1^\alpha(\bar{\gamma}_1+\bar{\gamma}_2)}{\bar{\gamma}_1\bar{\gamma}_2}\left(1-\frac{16}{\pi^2}\right)}+\bar{\gamma}_1 e^{\frac{d_2^\alpha(\bar{\gamma}_1+\bar{\gamma}_2)}{\bar{\gamma}_1\bar{\gamma}_2}\left(1-\frac{16}{\pi^2}\right)}}{2(\bar{\gamma}_1+\bar{\gamma}_2)}\\ \nonumber
		&+\theta_F\Bigg[\tfrac{\bar{\gamma}_2 e^{\frac{d_1^\alpha(\bar{\gamma}_1+\bar{\gamma}_2)}{\bar{\gamma}_1\bar{\gamma}_2}\left(1-\frac{16}{\pi^2}\right)}\left(1+e^{\frac{d_1^\alpha(\bar{\gamma}_1+\bar{\gamma}_2)}{\bar{\gamma}_1\bar{\gamma}_2}\left(1-\frac{16}{\pi^2}\right)}\right)
	     +\bar{\gamma}_1 e^{\frac{d_2^\alpha(\bar{\gamma}_1+\bar{\gamma}_2)}{\bar{\gamma}_1\bar{\gamma}_2}\left(1-\frac{16}{\pi^2}\right)}\left(1+e^{\frac{d_2^\alpha(\bar{\gamma}_1+\bar{\gamma}_2)}{\bar{\gamma}_1\bar{\gamma}_2}\left(1-\frac{16}{\pi^2}\right)}\right)}{2(\bar{\gamma}_1+\bar{\gamma}_2)}\\ 
		&-\tfrac{\bar{\gamma}_2 e^{\frac{d_1^\alpha(2\bar{\gamma}_1+\bar{\gamma}_2)}{\bar{\gamma}_1\bar{\gamma}_2}\left(1-\frac{16}{\pi^2}\right)}+\bar{\gamma}_1 e^{\frac{d_2^\alpha(2\bar{\gamma}_1+\bar{\gamma}_2)}{\bar{\gamma}_1\bar{\gamma}_2}\left(1-\frac{16}{\pi^2}\right)}}{(2\bar{\gamma}_1+\bar{\gamma}_2)}
		-\tfrac{2\bar{\gamma}_2 e^{\frac{d_1^\alpha(\bar{\gamma}_1+2\bar{\gamma}_2)}{\bar{\gamma}_1\bar{\gamma}_2}\left(1-\frac{16}{\pi^2}\right)}+\bar{\gamma}_1 e^{\frac{d_2^\alpha(\bar{\gamma}_1+2\bar{\gamma}_2)}{\bar{\gamma}_1\bar{\gamma}_2}\left(1-\frac{16}{\pi^2}\right)}}{2(\bar{\gamma}_1+2\bar{\gamma}_2).}
\Bigg]\Bigg)
		\end{align}
%	\hrulefill
\end{figure*}
\section{Coverage region}
By exploiting the concept of coverage region in \cite[Def. 2]{Aggarwal2009}, we determine the expression for the coverage region of the system model in Fig. \ref{system-model}. For simplicity and without loss of generality, we assume that receiver $r$ is located at the origin $(0,0)$. Then, we define the coverage region as the geographic area, i.e., the set of distances $d_1$ and $d_2$, for which the sum rate $R_1+ R_2$ is guaranteed, with $R_1, R_2>0$, i.e.,	\setcounter{equation}{22}
\begin{align}
\mathcal{G}(d_1,d_2)\overset{\text{def}}{=}\{d_1,d_2,\mathcal{C}(d_1,d_2)>R_1+R_2\}
\end{align}
where $\mathcal{C}(d_1,d_2)=\frac{1}{2}\log_2\left(1+\min\{\frac{P_1|h_1|^2}{Nd_1^\alpha},\frac{P_2|h_2|^2}{Nd_2^\alpha}\}\right)$ denotes the channel capacity when transmitters $t_1$ and $t_2$ are located at $d_1$ and $d_2$, respectively. We note that, while the definition of coverage region is related to an expectation over the achievable rate classically associated to an ergodic setting, it becomes relevant in a block-fading setting with link adaptation capabilities \cite{Lozano2012}. For the sake of compactness, the coverage region is exemplified for the FGM Copula.
	\setcounter{equation}{24}

\begin{theorem}\label{thm-cov}
The coverage region for the concerned correlated Rayleigh fading doubly dirty MAC with defined parameters $\bar{\gamma}_1$, $\bar{\gamma}_2$, $\theta_F$, $\alpha$, $R_1$, and $R_2$ is given by \eqref{coverage region}.%, where $\alpha>2$ is the path-loss exponent. \vspace{-2ex}
\begin{proof}
The coverage region is expressed in terms of an expectation over the random SNRs $\gamma_1$ and $\gamma_2$ as:
\begin{align}
&R_1+R_2\leq \mathbb{E}_{\gamma_1,\gamma_2}\Bigg[\frac{1}{2}\log_2\left(1+\min\left\{\frac{\gamma_1}{ {d}^{\alpha}_1},\frac{\gamma_2}{{d}^{\alpha}_2}\right\}\right)\Bigg]\\ 
&=\int_{0}^{\infty}\int_{0}^{\infty}\frac{1}{2}\log_2\left(1+\min\left\{\frac{\gamma_1}{ {d}^{\alpha}_1},\frac{\gamma_2}{{d}^{\alpha}_2}\right\}\right)f(\gamma_1,\gamma_2)d\gamma_1 d\gamma_2\\ \nonumber
&=\int_{0}^{\infty}\Bigg(\int_{0}^{\gamma_2}\frac{1}{2}\log_2\left(1+\frac{\gamma_1}{ {d_1^\alpha}}\right)f(\gamma_1,\gamma_2)d\gamma_1\\  
&\;\;\;+\int_{\gamma_2}^{\infty}\frac{1}{2}\log_2\left(1+\frac{\gamma_2}{ {d_2^\alpha}}\right)f(\gamma_1,\gamma_2)d\gamma_1\Bigg)d\gamma_2.\label{int-thm4}
\end{align}
where $f(\gamma_1,\gamma_2)$ is the joint PDF of the SNRs \cite{ghadi2020copula1}, obtained from Corollary \ref{col-pdf} by performing partial derivatives over the Copula function in \eqref{fgm} and using the chain rule, as follows
\begin{align}\label{pdf}
f(\gamma_1,\gamma_2)=\tfrac{e^{-\frac{\gamma_1}{\bar{\gamma}_1}-\frac{\gamma_2}{\bar{\gamma}_2}}}{\bar{\gamma}_1\bar{\gamma}_2}\left[1+\theta_F\left(1-2e^{-\frac{\gamma_1}{\bar{\gamma}_1}}\right)\left(1-2e^{-\frac{\gamma_2}{\bar{\gamma}_2}}\right)\right].
\end{align}
By substituting the joint PDF from \eqref{pdf} into \eqref{int-thm4}, and calculating the above integrals, the coverage region is obtained as \eqref{coverage region}. The details of the proof are in Appendix \ref{app1}.
\end{proof}
\end{theorem}
\vspace{-2mm}
\section{Numerical Results}
\begin{figure}[t]\vspace{0cm}
	\centering
	\includegraphics[height=5.5 cm,width=7.5cm]{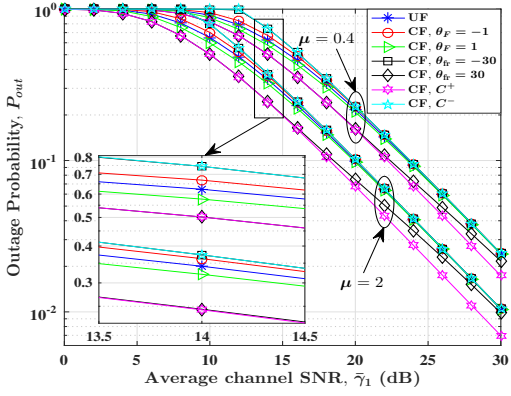} %\textwidth
	\caption{OP versus $\bar{\gamma}_1$ for selected values of dependence parameter $\theta_F$, $\theta_{\rm fr}$, and the ratio  $\mu=\bar{\gamma}_2/\bar{\gamma}_1$.}\vspace{0cm}
	\label{out-snr}
\end{figure}

In this section, the analytical and Monte-Carlo simulation results for the OP and coverage region are presented, with special focus on comparing the performances in the presence/absence of fading correlation. The path-loss exponent is $\alpha=3.5$ for all results. Fig. \ref{out-snr} shows the behavior of the OP based on the variation of $\bar{\gamma}_1$ for selected values of $\theta_F$, $\theta_{\rm fr}$, and $\mu=\bar{\gamma}_2/\bar{\gamma}_1$.
For simplicity, we set $d_1=d_2=1$ in this scenario. In all instances, the Fr\'echet-Hoeffding bounds provide the limit performance for the extreme cases of correlation. We see that under the positive dependence structure ($C^{+}, \theta_{\rm fr}=30,$ and $\theta_F=1$), the correlated fading (CF) case achieves better performance (lower OP) as compared to the uncorrelated fading (UF) case. 
\begin{figure}[t]%\vspace{-0.2cm}
	\centering
	\includegraphics[height=5.5 cm,width=7.5cm]{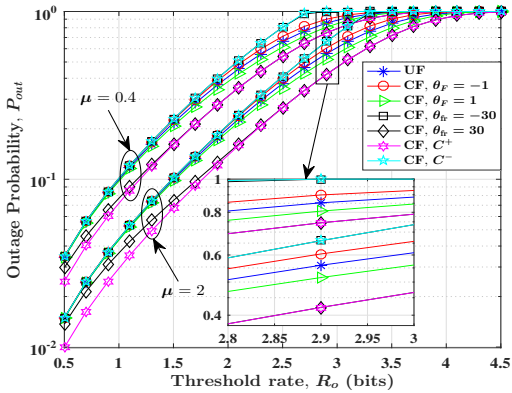} %\textwidth
	\caption{OP versus threshold rate $R_o$ for selected values of dependence parameter $\theta_F$, $\theta_{\rm fr}$, and the ratio $\mu=\bar{\gamma}_2/\bar{\gamma}_1$.}
	\label{out-r}
\end{figure}
The effect of the threshold rate $R_o$ on the OP for selected values of $\theta_F$, $\theta_{\rm fr}$ and $\mu$ is evaluated in Fig. \ref{out-r}. As $R_o$ increases, the OP tends to 1, which is coherent with the fact that communication becomes impossible at very high rates. We also confirm that positive dependence has a beneficial role on system performance (lower OP). 

\begin{figure}[t]%\vspace{-0.2cm}
	\centering
	\includegraphics[width=3in]{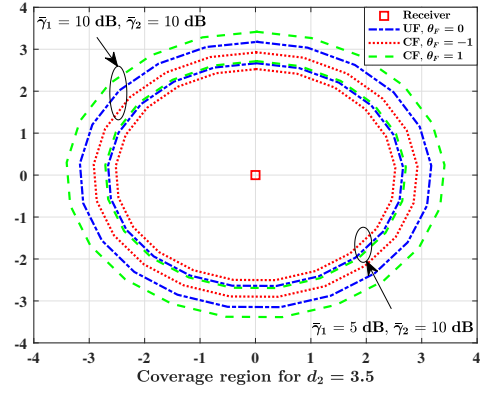} %\textwidth
	\caption{Coverage region for selected values of dependence parameter $\theta_F$.} 
	\label{cov-mac-side}
\end{figure}

The coverage region for selected values of $\theta_F$ and $\bar{\gamma}_1$ is illustrated in Fig. \ref{cov-mac-side}. We see that increasing $\bar{\gamma}_1$ extends the coverage region, and also that when $\bar{\gamma}_1$ and $\bar{\gamma}_2$ are similar the role of dependence becomes more noticeable. The bottleneck effect imposed by the user with a minimum SNR in the capacity region \eqref{capacity region}, is relaxed in the presence of a positive dependence. Thus, the coverage region is \textit{improved} compared to the case of independent fading, as observed in the figure. It is important to highlight that this is in stark contrast with the observations made in \cite{ghadi2020copula} in the absence of interference, for which the opposite conclusion was obtained. Hence, we see that considering the non-causally known SI at transmitters in MAC can improve the performance of OP and coverage region under the positive dependence structure.

\vspace{-2mm}
\section{Conclusion}
We evaluated the performance of doubly dirty MAC with non-causally known side information at transmitters, where the corresponding fading channel coefficients are assumed correlated. We derived closed-form expressions and lower/upper bounds for the OP, analyzing the effect of correlated fading under negative/positive dependence structures. We also characterized the coverage region using the FGM Copula, which covers negative and positive correlation. We showed that system performance is improved in terms of OP reduction and coverage region extension. Results confirm the beneficial impact of positive fading correlation in the doubly dirty MAC channel due to strong interference, compared to the case of an interference-free clean MAC.
\appendices
\vspace{-2mm}
\section{Proof of Theorem \ref{thm-cov}}\label{app1}\vspace{0cm}
After applying the joint PDF $f(\gamma_1,\gamma_2)$ in \eqref{int-thm4} and exploiting the linearity rules of integration, \eqref{int-thm4} can be decomposed as
\begin{align}\nonumber
R_1&+R_2\leq 
\int_{0}^{\infty}\int_{0}^{\gamma_2}\frac{e^{-\frac{\gamma_1}{\bar{\gamma}_1}-\frac{\gamma_2}{\bar{\gamma}_2}}}{2\bar{\gamma}_1\bar{\gamma}_2}\log_2\left(1+\frac{\gamma_1}{ d_1^{\alpha}}\right)\\ \nonumber
&\times\Big[1+\theta_F\left(1-2e^{-\frac{\gamma_1}{\bar{\gamma}_1}}\right)\left(1-2e^{-\frac{\gamma_2}{\bar{\gamma}_2}}\right)\Big]d\gamma_1 d\gamma_2\\\nonumber
&+\int_{0}^{\infty}\int_{\gamma_2}^{\infty}\frac{e^{-\frac{\gamma_1}{\bar{\gamma}_1}-\frac{\gamma_2}{\bar{\gamma}_2}}}{2\bar{\gamma}_1\bar{\gamma}_2}\log_2\left(1+\frac{\gamma_2}{d_2^{\alpha}}\right)\\
&\times\Big[1+\theta_F\left(1-2e^{-\frac{\gamma_1}{\bar{\gamma}_1}}\right)\left(1-2e^{-\frac{\gamma_2}{\bar{\gamma}_2}}\right)\Big]d\gamma_1 d\gamma_2\\\nonumber
&=\mathscr{A}_1+\theta_F(\mathscr{A}_1-2\mathscr{A}_2-2\mathscr{A}_3+4\mathscr{A}_4)\\
&+\mathscr{B}_1+\theta_F(\mathscr{B}_1-2\mathscr{B}_2-2\mathscr{B}_3+4\mathscr{B}_4).\label{int-5}
\end{align}
where the integrals in \eqref{int-5} follow the following formats:
\begin{align}\nonumber
&\int e^{-\zeta x}\log_2(1+\eta x)dx\\
&=\frac{1}{\zeta\ln 2}\Big[e^{\frac{\zeta}{\eta}}\mathrm{Ei}\left(-(\frac{\zeta}{\eta}+\zeta x)\right)-e^{-\zeta x}\ln(1+\eta x)\Big],\label{sol-int3}
\end{align}
\begin{align}
\int_{0}^{\infty}e^{-\zeta x}\log_2(1+\eta x)dx=-\frac{e^{\frac{\zeta}{\eta}}}{\zeta\ln 2}\mathrm{Ei}\left(-\frac{\zeta}{\eta}\right),\label{sol-int4}
\end{align}
\begin{align}\nonumber
&\int_{0}^{\infty}e^{-\zeta x}\mathrm{Ei}\left(-(\kappa+\eta x)\right)dx\\
&=\frac{1}{\zeta}\Big[\mathrm{Ei}(-\kappa)-e^{\frac{\zeta\kappa}{\eta}}\mathrm{Ei}\left(-\frac{(\zeta+\eta)\kappa}{\eta}\right)\Big].\label{sol-int5}
\end{align}
Now, by exploiting \eqref{sol-int3}, \eqref{sol-int4}, and \eqref{sol-int5}, we have:
\begin{align}\nonumber
\mathscr{A}_1&=\int_{0}^{\infty}\int_{0}^{\gamma_2}\frac{e^{-\frac{\gamma_1}{\bar{\gamma}_1}-\frac{\gamma_2}{\bar{\gamma}_2}}}{2\bar{\gamma}_1\bar{\gamma}_2}\log_2\left(1+\frac{\gamma_1}{d_1^{\alpha}}\right)d\gamma_1 d\gamma_2\\
&=-\frac{\bar{\gamma}_2 e^{d_1^\alpha(\frac{\bar{\gamma}_1+\bar{\gamma}_2}{\bar{\gamma}_1\bar{\gamma}_2})}}{2(\bar{\gamma}_1+\bar{\gamma}_2)\ln 2}\mathrm{Ei}\left(-d_1^\alpha\left(\frac{\bar{\gamma}_1+\bar{\gamma}_2}{\bar{\gamma}_1\bar{\gamma}_2}\right)\right),\label{a1}
\end{align}
\begin{align}\nonumber
\mathscr{A}_2&=\int_{0}^{\infty}\int_{0}^{\gamma_2}\frac{e^{-\frac{\gamma_1}{\bar{\gamma}_1}-\frac{2\gamma_2}{\bar{\gamma}_2}}}{2\bar{\gamma}_1\bar{\gamma}_2}\log_2\left(1+\frac{\gamma_1}{d_1^{\alpha}}\right)d\gamma_1d\gamma_2\\&=-\frac{\bar{\gamma}_2 e^{d_1^\alpha(\frac{2\bar{\gamma}_1+\bar{\gamma}_2}{\bar{\gamma}_1\bar{\gamma}_2})}}{2(2\bar{\gamma}_1+\bar{\gamma}_2)\ln 2}\mathrm{Ei}\left(-d_1^\alpha\left(\frac{2\bar{\gamma}_1+\bar{\gamma}_2}{\bar{\gamma}_1\bar{\gamma}_2}\right)\right),\label{a2}
\end{align}
\begin{align}\nonumber
\mathscr{A}_3&=\int_{0}^{\infty}\int_{0}^{\gamma_2}\frac{e^{-\frac{2\gamma_1}{\bar{\gamma}_1}-\frac{\gamma_2}{\bar{\gamma}_2}}}{2\bar{\gamma}_1\bar{\gamma}_2}\log_2\left(1+\frac{\gamma_1}{d_1^{\alpha}}\right)d\gamma_1d\gamma_2\\
&=-\frac{\bar{\gamma}_2 e^{d_1^\alpha(\frac{\bar{\gamma}_1+2\bar{\gamma}_2}{\bar{\gamma}_1\bar{\gamma}_2})}}{2(\bar{\gamma}_1+2\bar{\gamma}_2)\ln 2}\mathrm{Ei}\left(-d_1^\alpha\left(\frac{\bar{\gamma}_1+2\bar{\gamma}_2}{\bar{\gamma}_1\bar{\gamma}_2}\right)\right),\label{a3}
\end{align} 
\begin{align}\nonumber
\mathscr{A}_4&=\int_{0}^{\infty}\int_{0}^{\gamma_2}\frac{e^{-\frac{2\gamma_1}{\bar{\gamma}_1}-\frac{2\gamma_2}{\bar{\gamma}_2}}}{2\bar{\gamma}_1\bar{\gamma}_2}\log_2\left(1+\frac{\gamma_1}{d_1^{\alpha}}\right)d\gamma_1d\gamma_2\\ 
&=-\frac{\bar{\gamma}_2 e^{2d_1^\alpha(\frac{\bar{\gamma}_1+\bar{\gamma}_2}{\bar{\gamma}_1\bar{\gamma}_2})}}{8(\bar{\gamma}_1+\bar{\gamma}_2)\ln 2}\mathrm{Ei}\left(-2d_1^\alpha\left(\frac{\bar{\gamma}_1+\bar{\gamma}_2}{\bar{\gamma}_1\bar{\gamma}_2}\right)\right).\label{a4}
\end{align}
Similarly, by utilizing \eqref{sol-int4}, we have:
\begin{align}\nonumber
\mathscr{B}_1&=\frac{1}{2\bar{\gamma}_1\bar{\gamma}_2}\int_{0}^{\infty}\int_{\gamma_2}^{\infty}e^{-\frac{\gamma_1}{\bar{\gamma}_1}-\frac{\gamma_2}{\bar{\gamma}_2}}\log_2\left(1+\frac{\gamma_2}{d_2^{\alpha}}\right)d\gamma_1d\gamma_2\\
&=-\frac{\bar{\gamma}_1 e^{\frac{d_2^\alpha(\bar{\gamma}_1+\bar{\gamma}_2)}{\bar{\gamma}_1\bar{\gamma}_2}}}{2 (\bar{\gamma}_1+\bar{\gamma}_2)\ln 2}\mathrm{Ei}\left(-\frac{d_2^\alpha(\bar{\gamma}_1+\bar{\gamma}_2)}{\gamma_1\gamma_2}\right),\label{b1}
\end{align}
\begin{align}\nonumber
\mathscr{B}_2&=\frac{1}{2\bar{\gamma}_1\bar{\gamma}_2}\int_{0}^{\infty}\int_{\gamma_2}^{\infty}e^{-\frac{\gamma_1}{\bar{\gamma}_1}-\frac{2\gamma_2}{\bar{\gamma}_2}}\log_2\left(1+\frac{\gamma_2}{d_2^{\alpha}}\right)d\gamma_1d\gamma_2\\
&=-\frac{\bar{\gamma}_1 e^{\frac{d_2^\alpha(2\bar{\gamma}_1+\bar{\gamma}_2)}{\bar{\gamma}_1\bar{\gamma}_2}}}{2 (2\bar{\gamma}_1+\bar{\gamma}_2)\ln 2}\mathrm{Ei}\left(-\frac{d_2^\alpha(2\bar{\gamma}_1+\bar{\gamma}_2)}{\gamma_1\gamma_2}\right),\label{b2}
\end{align}
\begin{align}\nonumber
\mathscr{B}_3&=\frac{1}{2\bar{\gamma}_1\bar{\gamma}_2}\int_{0}^{\infty}\int_{\gamma_2}^{\infty}e^{-\frac{2\gamma_1}{\bar{\gamma}_1}-\frac{\gamma_2}{\bar{\gamma}_2}}\log_2\left(1+\frac{\gamma_2}{d_2^{\alpha}}\right)d\gamma_1d\gamma_2\\
&=-\frac{\bar{\gamma}_1 e^{\frac{d_2^\alpha(\bar{\gamma}_1+2\bar{\gamma}_2)}{\bar{\gamma}_1\bar{\gamma}_2}}}{4 (\bar{\gamma}_1+2\bar{\gamma}_2)\ln 2}\mathrm{Ei}\left(-\frac{d_2^\alpha(\bar{\gamma}_1+\bar{2\gamma}_2)}{\gamma_1\gamma_2}\right),\label{b3}
\end{align}
\begin{align}\nonumber
\mathscr{B}_4&=\frac{1}{2\bar{\gamma}_1\bar{\gamma}_2}\int_{0}^{\infty}\int_{\gamma_2}^{\infty}e^{-\frac{2\gamma_1}{\bar{\gamma}_1}-\frac{2\gamma_2}{\bar{\gamma}_2}}\log_2\left(1+\frac{\gamma_2}{d_2^{\alpha}}\right)d\gamma_1d\gamma_2\\
&=-\frac{\bar{\gamma}_1 e^{\frac{2d_2^\alpha(\bar{\gamma}_1+\bar{\gamma}_2)}{\bar{\gamma}_1\bar{\gamma}_2}}}{8 (\bar{\gamma}_1+\bar{\gamma}_2)\ln 2}\mathrm{Ei}\left(-\frac{2d_2^\alpha(\bar{\gamma}_1+\bar{\gamma}_2)}{\gamma_1\gamma_2}\right).\label{b4}
\end{align} 
These integrals allow us to express (17) in closed-form by inserting \eqref{a1}-\eqref{b4} into \eqref{int-5}. A very tight approximation across the entire range of arguments of $\mathrm{Ei}(\cdot)$ can be derived by using $\mathrm{Ei}(-x)\sim-\frac{\sqrt{\pi}}{2}e^{-(\frac{16}{{\pi}^2})x}$ \cite{alkheir2013accurate}. This completes the proof.
\bibliographystyle{IEEEtran}
\bibliography{sample.bib}
\end{document}